\documentclass[aip,reprint]{revtex4-1}

\usepackage{amssymb}
\usepackage{graphicx}
\usepackage{bm}

\draft 

\begin{document}


\title{Kinetic Flux Ropes: Bernstein-Greene-Kruskal Modes for the Vlasov-Poisson-Amp\`{e}re System} 



\author{C. S. Ng}
\email[]{cng2@alaska.edu}
\homepage[]{https://sites.google.com/a/alaska.edu/chungsangng}
\affiliation{Geophysical Institute, University of Alaska Fairbanks, Fairbanks, Alaska 99775, USA}


\date{\today}

\begin{abstract}
Electrostatic structures have been observed in many regions of space plasmas, including the solar wind, the magnetosphere, the auroral acceleration region. One possible theoretical description of some of these structures is the concept of Bernstein-Greene-Kruskal (BGK) modes, which are exact nonlinear steady-state solutions of the Vlasov-Poisson system of equations in collisionless kinetic theory. We generalize exact solutions of two-dimensional BGK modes in a magnetized plasma with finite magnetic field strength [Ng, Bhattacharjee, and Skiff, Phys. Plasmas {\bf13}, 055903 (2006)] to cases with azimuthal magnetic fields so that these structures carry electric current as well as steady electric and magnetic fields. Such nonlinear solutions now satisfy exactly the Vlasov-Poisson-Amp\`{e}re system of equations. Explicit examples with either positive or negative electric potential structure are provided.
\end{abstract}

\pacs{52.35.Sb, 52.25.Dg, 52.35.Mw, 52.25.Xz}

\keywords{BGK modes, Solitons, Plasma kinetic equations, Nonlinear phenomena, Magnetized plasmas}

\maketitle 

\section{INTRODUCTION}

High temperature plasmas relevant to fusion experiments, 
  space and astrophysics can be considered as collisionless 
  due to small collision frequency.\cite{RevModPhys.71.S404} 
Consequently,
  particle distributions in a collisionless plasma often deviate from Maxwellian.
  \cite{doi:10.1029/JA073i009p02839, doi:10.1063/1.1667501, doi:10.1029/2009JA014352}
Studies of the collisionless Vlasov equation have produced many important insights in kinetic theory, 
  e.g., Landau damping of linear plasma waves,
  \cite{Landau:1946jc, PhysRevLett.83.1974, PhysRevLett.92.065002, doi:10.1063/1.4789882} 
  and the existence of exact self-consistent steady-state nonlinear solutions 
  of the Vlasov-Poisson equations known as 
  Bernstein-Greene-Kruskal (BGK)  modes in one dimension (1D),
  \cite{PhysRev.108.546}
  i.e., planer structures. 
A large number of papers have been written on the subject of BGK modes
  (too many to be cited here but please refer to an interesting 
  recent review\cite{doi:10.1063/1.4976854}
  ),
  but the vast majority of these works were still within the original 1D framework. 

The method of constructing a 1D BGK mode is straightforward, 
  either by specifying the form of the distribution first or the electric potential first, 
  as found in some plasma physics textbooks.
  \cite{nicholson1983introduction, swanson2003plasma, gurnett2017introduction}
However, there has also been considerable interest in BGK modes in higher dimensions,
  i.e., 2D (long tube structures) or 3D (structures localized in all three dimensions). 
This is mainly due to the fact that 3D features of solitary wave structures in space-based observations 
  that cannot be explained by 1D BGK modes.
  \cite{PhysRevLett.81.826, doi:10.1029/98GL50870, doi:10.1029/2005JA011095, doi:10.1029/1998GL900304}
For example, 
  Ref.~\onlinecite{PhysRevLett.81.826} 
  shows that electrostatic solitary waves observed in the auroral ionosphere 
  have electric field components perpendicular to the background magnetic field 
  with comparable magnitude to the parallel component. 
This is inconsistent with a 1D BGK-like potential, 
  which only has  the parallel component, 
  but consistent with the structure of a single-humped solitary potential 
  that travels past the spacecraft along the magnetic field. 
Also, there are results from numerical simulations suggesting 
  higher dimensional BGK modes.
  \cite{PhysRevLett.83.2344, doi:10.1029/1999GL003709, doi:10.1029/2000GL003766} 

One kind of approximate 3D BGK modes has been constructed under the assumption of 
  an infinitely strong background magnetic field.
  \cite{2002PhDT........10C, doi:10.1029/2001GL013385, PhysRevE.69.055401}
The strong field assumption constrains charged particles to move along magnetic field lines, 
  and thus effectively reduces the problem to 1D. 
To demonstrate in principle that higher dimensional exact BGK modes can indeed exist, 
  we have constructed exact 3D BGK modes 
  (in the geometry of a localized spherically symmetric potential) 
  by allowing the electron distribution function to depend on the angular momentum 
  for an unmagnetized plasma.\cite{PhysRevLett.95.245004} 
Another kind of solutions that has greater relevance to space plasmas with finite magnetic field 
  has also been constructed under the 2D geometry with 
  cylindrical symmetry.\cite{doi:10.1063/1.2186187} 
Such solutions satisfy exactly the Vlasov-Poisson-Amp\`{e}re system of equations 
  with the distribution depending on both the energy and the canonical angular momentum.  
Indeed a Particle-in-Cell (PIC) simulation has found structures similar to our 2D exact solution.
  \cite{doi:10.1029/2008JA013693} 
To study the stability of such 2D BGK modes, 
  we have also performed 
  2D PIC simulations,\cite{doi:10.1063/1.4723590} 
  which show that they are indeed stable for stronger magnetic field. 
  
Electrostatic structures have long been observed experimentally 
  in both magnetized plasmas,
  \cite{Lynov_1979,      PhysRevLett.42.501} 
  trapped pure electron plasmas,\cite{PhysRevLett.92.245003} 
  and in space-based observations of solitary waves or phase space holes
  in the magnetosphere and the solar wind.
  \cite{PhysRevLett.48.1175, PhysRevLett.61.82, M_lkki_1989,          doi:10.1029/94GL01284, doi:10.1029/97JA00684, PhysRevLett.79.1281, doi:10.1029/98GL02111, doi:10.1029/1999GL900207}
Early numerical simulations have also shown that 1D BGK modes can be formed dynamically 
  via a two-stream instability or nonlinear Landau damping 
  and can be reasonably stable.
  \cite{PhysRevLett.19.297,  doi:10.1063/1.1692361, PhysRevLett.23.1087, doi:10.1063/1.1693039, doi:10.1063/1.863392,       doi:10.1063/1.866579, PhysRevLett.48.481, doi:10.1029/95JA03145, PhysRevLett.79.2815}

Meanwhile, 
  kinetic physics down to small electron scales has increasingly been recognized 
  to play a central role in the fundamental process of magnetic reconnection, 
  which has been extensively studied in the space physics, 
  plasma physics, 
  and astrophysics communities for decades.
  \cite{sweet_1958, park57,            1964NASSP..50..425P,  RevModPhys.58.183, RevModPhys.58.741, prief00,                           biskamp2000magnetic, doi:10.1146/annurev.astro.42.053102.134039, birn2007reconnection, RevModPhys.82.603, Karimabadi2014,                        Lazarian2015,           Hesse2016,           Goldstein2016}
One consequence of kinetic physics during reconnection is the formation of small kinetic scale structures, 
  as seen by many kinetic simulations of reconnection.
  \cite{Drake873,          doi:10.1063/1.2218817, doi:10.1029/2005JA011542, doi:10.1029/2010GL043608, doi:10.1063/1.3529365, PhysRevLett.106.065002, doi:10.1063/1.4766895, doi:10.1063/1.4704799, doi:10.1002/2014JA020054, PhysRevLett.112.145002}
While most kinetic simulations are still in 2D due to high computing costs, 
  larger 3D kinetic simulations using PIC method have been performed in recent years and 
  have found that some of these structures having the geometric form of 
  magnetic flux ropes in 3D.\cite{daughton2011role} 
Physically, 
  such flux ropes are simply 3D counterparts of plasmoids (or secondary islands) 
  seen in 2D reconnection simulations, 
  either kinetic,\cite{doi:10.1063/1.2218817} 
  or even in resistive magnetohydrodynamics (MHD).
  \cite{doi:10.1063/1.3264103,          2011ASPC..444..124N}
  
Besides observations of possible BGK modes mentioned above, 
  there have been more observations of small-scale kinetic structures 
  in the magnetosphere in the last few years, 
  \cite{doi:10.1029/2010JA015546, doi:10.1029/2012JA018141, doi:10.1029/2012JA017552, angeo-30-583-2012, doi:10.1002/2014JA019924, doi:10.1002/2014JA020856, doi:10.1002/2014JA020786, doi:10.1002/2016GL068545, doi:10.1002/2016GL069157}
  from spacecraft such as the Magnetospheric Multiscale (MMS) Mission. 

Motivated by the formation of kinetic flux ropes seen in 3D simulations, 
  we have generalized the form of exact 2D BGK mode solutions further, 
  now with the distribution function also depending on the 
  $z$-component (along the symmetric axis) 
  of the canonical momentum. 
Such more general solutions thus allow a parallel current density with
  associated azimuthal magnetic field.
The resulting structures are then of the form of magnetic flux ropes,
  due to kinetic physics,
  rather than MHD.
We present the analytic forms of such solution in Section~\ref{analytic}.
Some solutions will then be solved numerically in Section~\ref{numeric} 
  to show properties of solutions as examples.
Discussions and conclusion will be presented in Section~\ref{conclu}.

\section{CONSTRUCTION OF SOLUTIONS \label{analytic}}

We start the construction of BGK mode solutions by requiring the 
  $s$-charge species to satisfy the Vlasov Equation
\begin{equation}
  \frac{\partial f_s}{\partial t} + \textbf{v}\cdot\frac{\partial f_s}{\partial \textbf{r}} 
  + \frac{q_s}{m_s}(\textbf{E}+\textbf{v}\times\textbf{B})\cdot
  \frac{\partial f_s}{\partial \textbf{v}} = 0 \; ,  \label{vlasov}
\end{equation}
  where $s = e$ for electrons, and $s = i$ for ions.
Note that $q_e = -e$ is a negative value for the electron charge,
  and $m_e$ is the mass of electron.
In this paper,
  we continue to follow the usual simplification of setting ions to be forming a
  uniform background of positive ion charge density $n_0 e$
  to maintain charge neutrality outside the localized
  BGK mode structure.\cite{2002PhDT........10C} 
One situation this approximation is justified is when
  the electrons, 
  imbedded with the BGK structure, 
  are drifting through a uniform background of ions with high enough
  relative velocity such that ions cannot respond significantly 
  due to the heavier ion mass during the short transit time.
This approximation is also valid in the limit of large ion to electron temperature ratio,
  even without a large relative drift between ions and electrons.
While this assumption seems drastic,
  it has been found that qualitatively similar solutions still
  exist for a large range of ion to electron temperature ratio,
  \cite{2018AGUFMSM13B2870N} 
  which will be the subject of a future publication.
By using the form of the Vlasov Equation in Eq.~(\ref{vlasov}),
  we have assumed a non-relativistic treatment.
This is because the numerical construction of BGK mode solutions
  for the relativistic Vlasov Equation is much more complicated.
For BGK modes,
  we seek steady-state solutions and thus we ignore the time derivative term.
We also seek self-consistent solutions so that the electric field {\bf E} 
  and the magnetic field {\bf B} in Eq.~(\ref{vlasov}) are solved from the Gauss Law
  $\nabla \cdot \textbf{E} = \rho_q/\epsilon_{0}$ and the 
  steady-state Amp\`{e}re Law
  $ \nabla\times\mathbf{B}=\mu_{0}\mathbf{J}$,
  with charge density $\rho_q$ and current density {\bf J} 
  calculated from integrating the moments of the distribution function $f_e$.
Introducing electric potential $\psi$ and magnetic potential {\bf A} such that
  $\textbf{E} = -\nabla \psi$ and $\textbf{B} = \nabla \times \textbf{A}$,
  the Gauss Law (Poisson equation)
  and the Amp\`{e}re Law become
\begin{eqnarray}
   \nabla^2 \psi & = &
    -\frac{e}{\varepsilon_0}\left[n_0 - \int d\textbf{v}f_e \right] \; , 
     \label{gauss} 
     \\
   \nabla \times \nabla \times \textbf{A} & = &
    -e \mu_0 \int d\textbf{v} f_e \textbf{v} \; ,
     \label{ampere}
\end{eqnarray}
  where we have ignored possible current density 
  due to the background ions, 
  assuming the drift velocity between electrons and ions
  is much smaller than the electron thermal velocity.

Since we are looking for 2D cylindrically symmetric solutions,
  it is convenient to use cylindrical coordinates,
  with the $z$-axis being the symmetric axis.
Therefore,
  only the radial coordinate $\rho$ enters into the spatial
  dependence of all physical quantities.
The steady-state Vlasov Equation can then be written explicitly as
\begin{eqnarray}
 &&  v_\rho\frac{\partial f_e}{\partial \rho}
   +   \left\{\frac{e}{m_e}\left[\frac{d\psi}{d\rho}-\frac{v_\phi}{\rho}
  \frac{d(\rho A_\phi)}{d\rho}-v_z\frac{dA_z}{d\rho}\right]
  +\frac{v_\phi^2}{\rho}\right\}\frac{\partial f_e}{\partial v_\rho} 
  \nonumber  \\
 &&  -  \left[\frac{v_\rho v_\phi}{\rho}  -  \frac{ev_\rho}{m_e}
  \frac{d(\rho A_\phi)}{\rho d\rho}\right]\frac{\partial f_e}{\partial v_\phi} 
  + \frac{ev_\rho}{m_e}\frac{dA_z}{d\rho}\frac{\partial f_e}{\partial v_z}=0 
  \; ,
  \label{vlasov-cyl} 
\end{eqnarray}
  with $\textbf{A}  = A_\phi \hat{\phi} + A_z \hat{z}$
  and $\textbf{v}  = v_\rho \hat{\rho} + v_\phi \hat{\phi} + v_z \hat{z}$.
Before we continue the construction of solutions for this equation,
  we introduce the following units to normalize physical quantities
  to shorten expressions in analysis,
  as well as in numerical codes.
We will measure velocity {\bf v} in the unit of electron thermal velocity
  $v_e$ of the background (far away from the electrostatic structure) 
  Maxwellian electrons,
  spatial vector {\bf r} in the unit of the Debye length
  $\lambda = v_e/\omega_{pe} = (v_e/e)\sqrt{\varepsilon_0 m_e/n_0}$
  (with $\omega_{pe}$ being the electron plasma frequency),
  electric potential $\psi$ in the unit of
  $n_0 e\lambda^2/\varepsilon_0$,
  electron distribution function $f_e$ in the unit of
  $n_0/v_e^3$,
  and magnetic field {\bf B} in the unit of
  $n_0 e\lambda/\varepsilon_0 v_e$,
  with units of other quantities derived from combinations
  of these units.
Note that a magnetic field with a field strength of unity
  ($B=1$) in this unit indicates a case where
  the electron cyclotron frequency equal to $\omega_{pe}$.
The Vlasov equation in these units is simply given by
  Eq.~(\ref{vlasov-cyl}) without the need of writing out the
  $e/m_e$ factors.

For the 2D case with cylindrical symmetry,
  the steady-state Vlasov equation can be solved
  by a $f_e$ that depends only on conserved quantities.
Therefore, 
  we seek solutions of the general form of 
  $f_e = f(w, l, p)$,
  where $w = v^2/2 - \psi = (v^2_\rho + v^2_\phi + v^2_z)/2 - \psi$
  is the normalized total (kinetic plus electrostatic) energy of a particle,
  $l = 2 \rho (v_\phi - A_\phi)$ 
  is two times the $z$-component of the normalized canonical angular momentum,
  and $p = v_z - A_z$ 
  is the $z$-component of the normalized canonical momentum.
Note that solutions in Ref.~\onlinecite{doi:10.1063/1.2186187} 
  is for the special case without the $p$ dependence.
This generalization turns out to be non-trivial since the $p$ dependence will
  generally induce a current along the $z$-axis which in turn
  produces an azimuthal magnetic field and thus the 
  flux rope structure.
So this should be the more general case in
  physical situations.
To complete the solution,
  $f_e$ of this form must also satisfies self-consistently
   the Gauss Law (\ref{gauss}) 
   and the Amp\`{e}re Law (\ref{ampere}),
   which can be written explicitly as
\begin{eqnarray}
  \frac{1}{\rho}\frac{d}{d\rho}\left(\rho\frac{d\psi}{d\rho}\right)
  & = & \int d^3 v f(w,l,p) -1 \; ,
     \label{gauss-rho} 
  \\
  \frac{d}{d\rho}\left[\frac{1}{\rho}\frac{d}{d\rho}\left(\rho A_\phi \right)\right] 
  & = & \beta^2_e \int d^3 v f(w,l,p) v_\phi  \; ,
     \label{ampere-phi}
  \\
  \frac{1}{\rho}\frac{d}{d\rho}\left(\rho \frac{d A_z}{d\rho} \right)
  & = & \beta^2_e \int d^3 v f(w,l,p) v_z  \; ,
     \label{ampere-z}
\end{eqnarray}
  where $\beta^2_e = v^2_e/c^2$ 
  is the square of the ratio of the electron thermal velocity
  to the speed of light.
These three equations are much more complicated 
  than simply three coupled ordinary differential equations,
  as their forms seem to suggest.
This is due to the fact that $f$ is an unknown
  function that depends on $\psi$, $A_\phi$, and $A_z$.
Therefore this is a set of three coupled nonlinear
  integral-differential equations,
  with the possibility of having many non-trivial solutions.
The fact that we are using a non-relativistic treatment
  might suggest simply setting $\beta^2_e = 0$,
  so that Eqs.~(\ref{ampere-phi}) and (\ref{ampere-z})
  imply special solutions of
  $A_\phi = B_{z0}\rho/2$, 
  $A_z = {\rm constant}$,
  or simply a uniform magnetic field along the $z$-direction
  with a field strength of $B_{z0}$.
This limit is essentially considering the electrostatic
  effect only,
  which is indeed the dominant effect.
However,
  for solutions having flux rope structures,
  we require the self-generated non-uniform
  magnetic field.
Therefore we will consider a large enough $\beta^2_e$
  but still within the validity of the non-relativistic treatment.

Due to the (maybe infinitely) large number of acceptable
  forms of the function $f$,
  we will first show the existent of solutions by limiting it to a form of
\begin{equation}
  f(w,l,p) = (2\pi)^{-3/2}e^{-w}\left(1-h e^{-k l^2 -\xi p^2} \right)
   \; ,  \label{f-form}
\end{equation}
  where $-\infty < h < 0$ or $ 0 < h < 1$, $k > 0$, $\xi > 0$ are
  constant parameters.
The case of $h = 0$ is excluded explicitly due to the 
  non-existence of localized solutions as proved in 
  Ref.~\onlinecite{PhysRevLett.95.245004}. 
This form with the specified ranges of parameters
  is chosen to ensure $\infty > f \geq 0$ 
  for any $\rho$ and {\bf v} values,
  since physically it is a distribution function,
  and that $f$ tends to a Maxwellian (with the 
  already specified thermal velocity) as $\rho \rightarrow \infty$
  (assuming $\psi \rightarrow 0$,  
  and the magnetic field tends to a uniform field,
  in the same limit)
  so that the solution represents a localized structure
  in the transverse directions.
Moreover, 
  this form of $f$ would allow the integration of its 
  moments easier,
  and resulting in analytic expressions 
  in the right-hand-sides of Eqs.~(\ref{gauss-rho}) to (\ref{ampere-z}).
While this is one of many possible forms,
  it already have three parameters that allows solutions
  to have a large range of properties that require
  an extensive scan of the parameter space to fully explore.
Obvious generalizations of this form include
  adding terms with different $h$, $k$, and $\xi$,
  or adding terms with more general dependence in $l$ or $p$ 
  in the exponential function.
However,
  due to the nonlinear nature of this set of equations,
  it is unlikely to expand $f$ with a complete set of functions.
We will therefore only consider the form given by Eq.~(\ref{f-form}) in
  the rest of this paper.

Once we have specified the form of $f$ using Eq.~(\ref{f-form}),
  we can then evaluate its moments to be used in 
  Eqs.~(\ref{gauss-rho}) to (\ref{ampere-z}).
The zeroth order moment gives the normalized electron density
  $n_e = \int d^3 v f(w,l,p)$,
  with an analytic expression after integration
\begin{equation}
  n_e = e^\psi \left[1-\frac{h\exp\left(-\frac{4 k A_\phi^2 \rho^2}{1+8 k \rho^2}
  -\frac{\xi A_z^2}{1+2 \xi} \right)}
  {\sqrt{\left(1+8k \rho^2\right)\left(1+2 \xi \right)}}
  \right]
   \; ,  \label{ne}
\end{equation}
  which can be substituted as the first term on the right-hand-side of 
  Eq.~(\ref{gauss-rho}).
Note that $n_e$ is always positive with the restrictions on parameters
  mentioned above.
Also,
  $n_e \rightarrow 1$ as $\rho \rightarrow \infty$,
  indicating a localized structure as required.
The expression in Eq.~(\ref{ne}) gives further restrictions
  for the existence of a localized $\psi$,
\begin{equation}
  1 \gtrless e^{\psi(0)} \left[1-\frac{h}{\sqrt{
  1+2 \xi }}
  \exp\left(-\frac{\xi A_z^2(0)}{1+2 \xi} \right)\right]
   \; ,  \label{h-limit}
\end{equation}
  with $\psi(0)$ and $A_z(0)$ being $\psi$ and $A_z$ at $\rho = 0$,
  for positive/negative $h$ and $\psi(0)$.

Similarly, 
  the first order moments give the normalized current
  density $\textbf{J}  = J_\phi \hat{\phi} + J_z \hat{z}$,
  where $J_\phi = -\int d^3 v f(w,l,p) v_\phi$,
  and $J_z = -\int d^3 v f(w,l,p) v_z$.
The normalized electron flow velocity is then given by
  $\bar{\textbf{v}} = -\textbf{J}/n_e$.
In the same way,
  $J_\phi$ and $J_z$ can be integrated into analytic forms
\begin{eqnarray}
  J_\phi 
  & = & \frac{8 h k A_\phi  \rho^2 e^\psi}{\left(1+8 k \rho^2 \right)^{3/2} \sqrt{1+2 \xi}}
  \exp \left(-\frac{4 k A_\phi^2 \rho^2}{1+8 k \rho^2}
  -\frac{\xi A_z^2}{1+2 \xi}\right)  \; ,
     \nonumber
  \\
  J_z
  & = &  \frac{2 h \xi A_z e^\psi}{\sqrt{1+8 k \rho^2}\left(1+2 \xi \right)^{3/2}}
  \exp \left(-\frac{4 k A_\phi^2 \rho^2}{1+8 k \rho^2}
  -\frac{\xi A_z^2}{1+2 \xi}\right)  \; ,
     \nonumber
\end{eqnarray}
  which can then be substituted into Eqs.~(\ref{ampere-phi}) 
  and (\ref{ampere-z}) respectively.
With these analytic integrations,
  the right-hand-sides of Eqs.~(\ref{gauss-rho}) to (\ref{ampere-z})
  become explicit functions of $\psi$, $A_\phi$, and $A_z$
  such that they are a set of three coupled
  nonlinear ordinary differential equations
  that can be integrated numerically with a technique 
  described in Refs.~\onlinecite{PhysRevLett.95.245004} 
  and \onlinecite{doi:10.1063/1.2186187}. 
  
The second or higher order moments are not required in solving 
  for solutions.
However,
  for completeness and possible comparison with either 
  observation data or direct simulations,
  we will calculate the second order moments in terms of
  the normalized electron pressure tensor 
   $\bm{P} = \int d^3 v f(w,l,p) ({\bf v}-\bar{\textbf{v}})({\bf v}-\bar{\textbf{v}})$,
   which is in the unit of $n_0 k_B T_{e0}$ with $T_{e0} \equiv m_e v_e^2/k_B$
   and $k_B$ being the Boltzmann constant.
Expressing {\bf \em P} in cylindrical coordinates,
  it is easy to show that $P_{\rho \phi} = P_{\phi \rho} =
  P_{\rho z} = P_{z \rho} = 0$,
  and $P_{\rho \rho} = n_e$.
Other components are given by
\begin{eqnarray}
  P_{\phi \phi} + n_e \bar{v}_\phi^2
  & = &  e^\psi \left\{1 - h\left[\frac{1 + 8 k \rho^2 +\left(8 k A_\phi \rho^2 \right)^2}
  {\left(1+8k \rho^2\right)^{5/2}\sqrt{1+2 \xi}}\right] \right.
  \nonumber \\
  &  & \left. \exp\left(-\frac{4 k A_\phi^2 \rho^2}{1+8 k \rho^2}
  -\frac{\xi A_z^2}{1+2 \xi} \right)\right\}
  \; ,  \nonumber 
  \\
  P_{zz} + n_e \bar{v}_z^2 
  & = &  e^\psi \left\{1 - h\left[\frac{1 + 2 \xi +\left(2 \xi A_z \right)^2}
  {\left(1+2 \xi \right)^{5/2}\sqrt{1+8k \rho^2}}\right] \right.
  \nonumber \\
  &  & \left. \exp\left(-\frac{4 k A_\phi^2 \rho^2}{1+8 k \rho^2}
  -\frac{\xi A_z^2}{1+2 \xi} \right)\right\}
  \; ,  \nonumber 
  \\
  P_{\phi z} + n_e \bar{v}_\phi \bar{v}_z
  & = &  - \frac{16 h k \xi A_\phi A_z \rho^2}
  {\left(1+2 \xi \right)^{3/2}\left(1+8k \rho^2\right)^{3/2}}
  \nonumber \\
  && \exp\left(\psi - \frac{4 k A_\phi^2 \rho^2}{1+8 k \rho^2}
  -\frac{\xi A_z^2}{1+2 \xi} \right)
  \; ,      \label{P-comps}
\end{eqnarray}
  with $P_{z \phi} = P_{\phi z}$.
Clearly as $\rho \rightarrow \infty$,
  $P_{\rho \rho} = P_{\phi \phi} = P_{zz} \rightarrow 1$,
  with other components being zero.
We can also define normalized parallel and 
  perpendicular electron temperature,
  in the unit of $ k_B T_{e0}$,
  with $T_{e \|} \equiv P_{zz}/n_e$,
  and $T_{e \perp} \equiv (P_{\rho \rho} + P_{\phi \phi})/2n_e$.
The total normalized electron temperature is then
  $T_e \equiv (T_{e \|} + 2 T_{e \perp})/3$.

\section{NUMERICAL EXAMPLES \label{numeric}}

In this section,
  we will present some numerical examples to show
  that solutions constructed by the method 
  described in Section~\ref{analytic} do exist.
We have mentioned already that this group of solutions
  depend on a number of parameters: 
  $h$, $k$, $\xi$, and $\beta_e$.
Moreover, 
  in addition to the boundary condition of 
  $\psi \rightarrow 0$ as $\rho \rightarrow \infty$
  required for a localized solution,  
  two more free parameters are needed 
  for boundary conditions.
In our method of integration,
  we specify $A_z(0)$ and $B_z(0) = 2 dA_\phi(0)/d\rho$
  on the symmetric axis as those boundary values.
Note that  $A_\phi(0)$, $d\psi (0)/d\rho$, and $dA_z(0)/d\rho$
  are zero by symmetry.
The value $\psi(0)$ is determined numerically to 
  satisfy the condition for a localized solution 
  as described in Refs.~\onlinecite{PhysRevLett.95.245004} 
  and \onlinecite{doi:10.1063/1.2186187}. 
Because of a large parameter space with these six parameters,
  it is difficult to investigate how properties of solutions
  depend on these parameters,
  although we have simulated a large number of cases
  with different parameters to confirm that solutions do
  exist generally.
In this section however,
  we will present only two examples for brevity,
  to demonstrate explicitly the existence of solutions,
  and to point out some of their properties.
The first example is for a positive electric potential $\psi$ with $h > 0$,
  while the other one is for a negative $\psi$ with $h < 0$.
  
\begin{figure}
  \centerline{\includegraphics[width=3.3in]{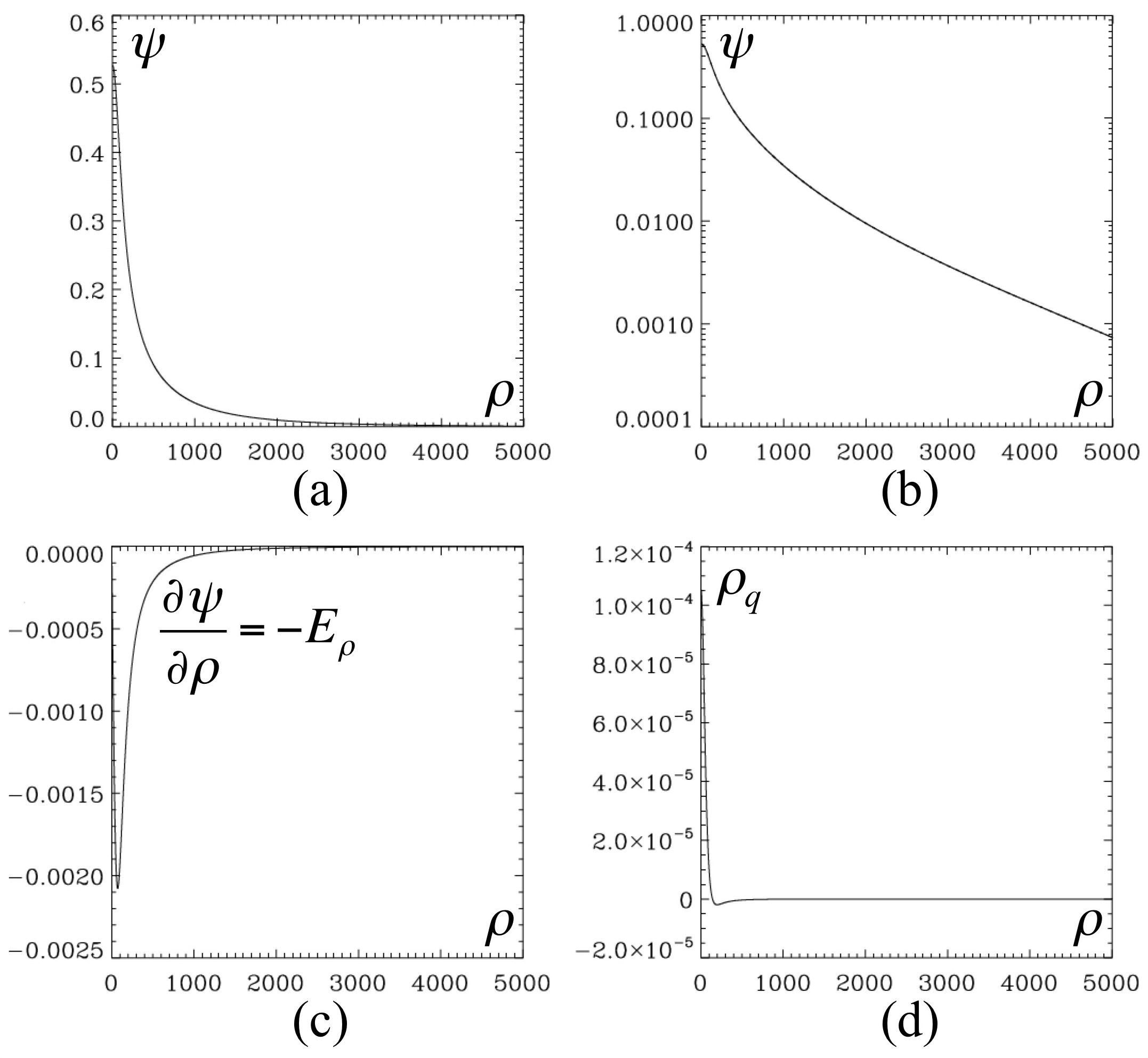}}
\caption{\label{fig1} (a) $\psi$, (b) $\psi$ in semi-log scales, 
  (c) $d\psi/d\rho$, (d) $\rho_q$ as functions of $\rho$, 
  for a case with $h = 0.99$, $k = 1 \times 10^{-5}$,
  $\xi = 1$, $\beta_e = 0.005$, $A_z(0) = 1$,
  and $B_z(0) = 0.00293$.}
\end{figure}

Fig.~\ref{fig1} shows plots of $\psi$, $d\psi/d\rho$, 
  which is the negative of the normalized radial electric field $E_\rho$,
  and the normalized charge density $\rho_q$ 
  as functions of the radial coordinate $\rho$,
  for the first case with $h = 0.99$, $k = 1 \times 10^{-5}$,
  $\xi = 1$, $\beta_e = 0.005$, $A_z(0) = 1$,
  and $B_z(0) = 0.00293$.
We have plotted over a range of $\rho$ from 0 to 5000,
  in the unit of $\lambda_D$,
  to show clearly the structures of the solution well into
  the asymptotic regime where $\psi \rightarrow 0$.
The length scale $l_\psi$ over which $\psi > 0.1 \psi(0)$ is 
  about $\rho < l_\psi \sim 700$,
  while the length scale $l_q$ over which $\rho_q > 0.1 \rho_q(0)$ is 
  about $\rho < l_q \sim 100$.
This example shows that the electrostatic structure
  and the region of charge non-neutrality can be much greater
  than $\lambda_D$,
  although there are also many other choices of parameters
  in which length scales
  are of the order of $\lambda_D$.
Another spatial unit that can be used to measure such kinetic
  structures is the electron inertia length 
  $d_e = c/\omega_{pe} = \lambda_D/\beta_e$.
For this case,
  we have $d_e = 200 \lambda_D$
  so that the range of $\rho$ is 0 to $25 d_e$ in Fig.~\ref{fig1},
  and $l_\psi \sim 3.5 d_e$, $l_q \sim 0.5 d_e$.
Note that $l_q \ll l_\psi$ is expected since the electric potential
  is due to the charge non-neutrality through the Possion equation.
Fig.~\ref{fig1}~(b) shows the plot of $\psi$ again in semi-log
  scales to show better the asymptotic behavior of
  $\psi$ as $\rho \rightarrow \infty$.
In fact,
  from Eq.~(\ref{gauss-rho}) with Eq.~(\ref{ne}),
  it can be shown that 
  $\psi \propto \exp(-\rho)/\sqrt{\rho}$
  in this limit.
From Fig.~\ref{fig1}~(d) we see that $\rho_q$
  becomes negative outside the positive core
  with a radius about $l_q$.
In fact the whole structure must be charge neutral,
  as required by the asymptotic behavior of the
  electric field which tends to zero exponentially
  at large $\rho$.
Fig.~\ref{fig1}~(c) shows that the positive radial
  electric field $E_\rho$ peaks around the edge
  of the core with positive charge density
  around $\rho \sim l_q$,
  and tends to zero over a wider tail of the
  order of $l_\psi$.
For this case, 
  the values of $\rho_q$ is very small,
  with a maximum about $10^{-4}$.
However, such a small $\rho_q$ can still produce
  a significant $\psi$ since it is non zero over a larger
  range of $\rho$.
With such a small $\rho_q$,
  the normalized electron density $n_e = 1 - \rho_q$
  is very close to unity.
Therefore we do not show a plot of $n_e$ here.

\begin{figure}
  \centerline{\includegraphics[width=3.3in]{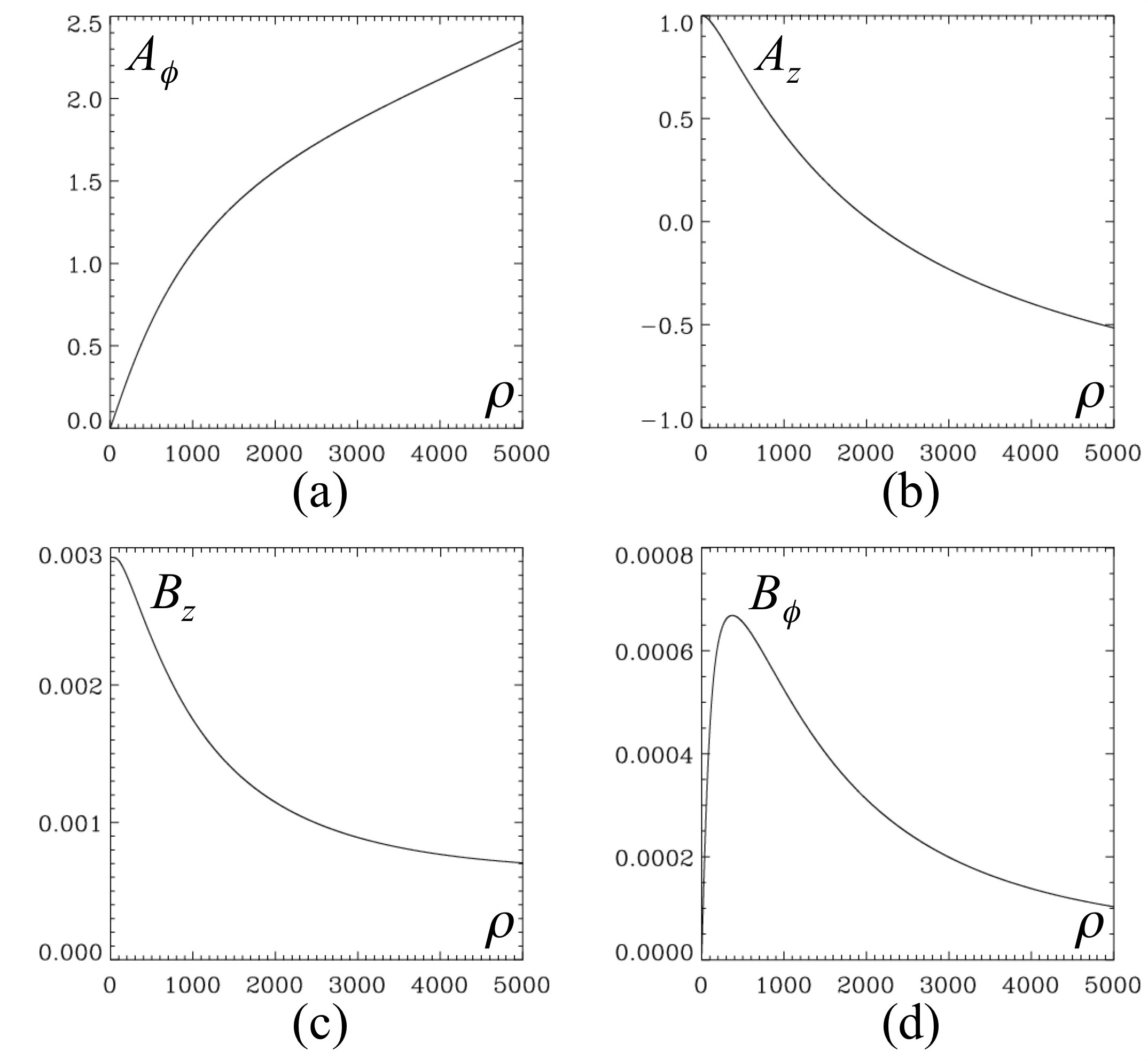}}
\caption{\label{fig2} (a) $A_\phi$, (b) $A_z$, 
  (c) $B_z$, (d) $B_\phi$ as functions of $\rho$, 
  for the same case as Fig.~\ref{fig1}.}
\end{figure}

Fig.~\ref{fig2} shows plots of magnetic potential components
  $A_\phi$, $A_z$ and corresponding magnetic field
  components $B_z = [d (\rho A_\phi)/d\rho]/\rho$,
  $B_\phi = -dA_z/d\rho$ for the same case.
Asymptotically as $\rho \rightarrow \infty$,
  $A_\phi \rightarrow B_\infty \rho/2$
  and $A_z \rightarrow -C_1 \ln \rho + C_2$ with
  $B_\infty$, $C_1$ and $C_2$ being constants,
  so that
  $B_z \rightarrow B_\infty$, 
  and $B_\phi \rightarrow C_1/\rho$,
  i.e.,
  a uniform field along the $z$-direction.
We can see that the length scale of the magnetic field structure 
  is much larger than the length scale of the electric
  structure $l_\psi$ and charge non-neutrality structure $l_q$.
This is expected since a plasma has a tendency to neutralize
  charge separation and to shield electric field,
  but does not neutralize current density and 
  does not shield magnetic field produced by net currents.
 
\begin{figure}
  \centerline{\includegraphics[width=3.3in]{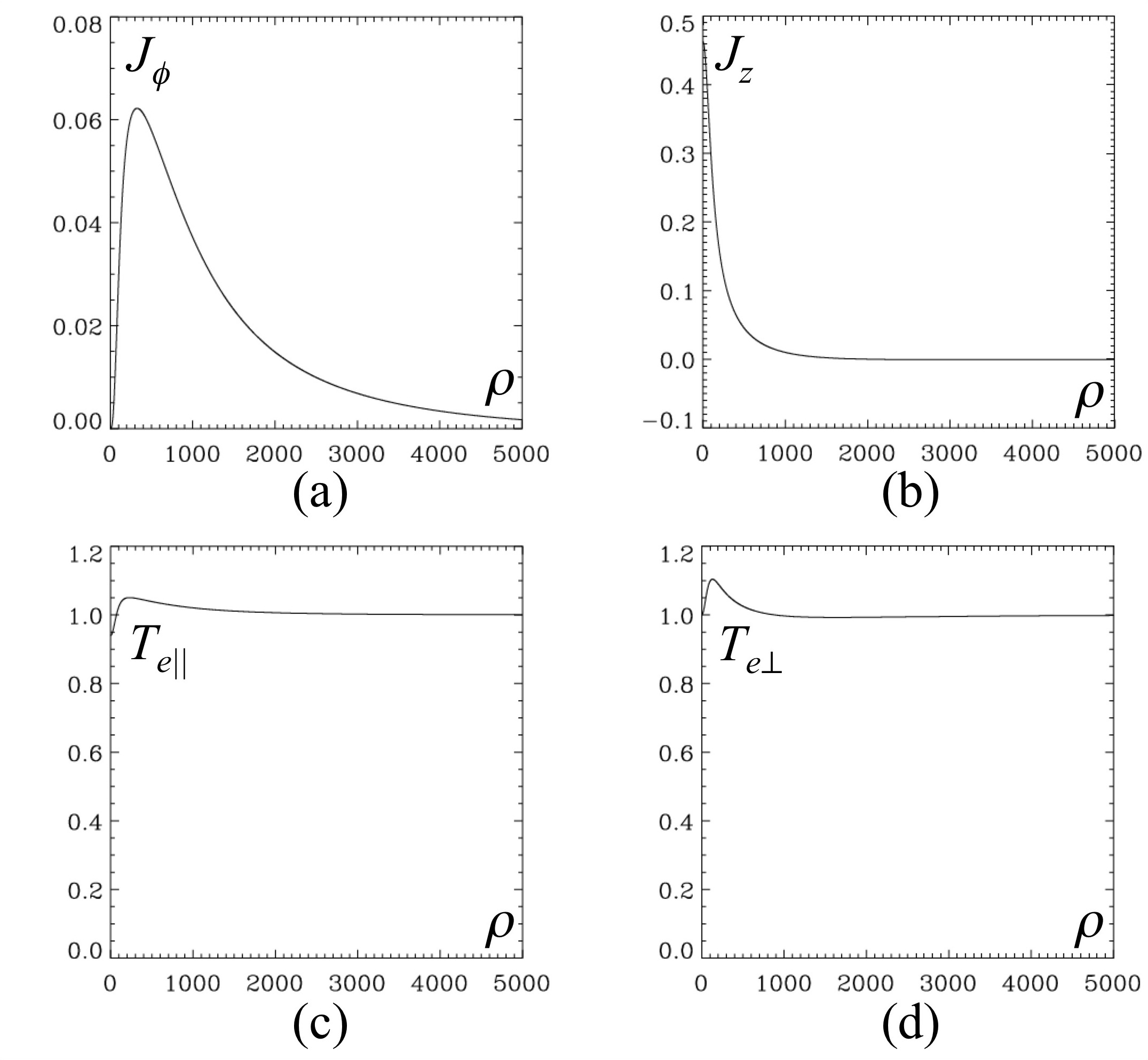}}
\caption{\label{fig3} (a) $J_\phi$, (b) $J_z$, 
  (c) $T_{e\parallel}$, (d) $T_{e\perp}$ as functions of $\rho$, 
  for the same case as Fig.~\ref{fig1}.}
\end{figure}

Fig.~\ref{fig3} (a) and (b) shows plots of the normalized current density 
  components $J_\phi$ and $J_z$,
  corresponding to the magnetic field shown in Fig.~\ref{fig2}~(c) and (d),
  through the Amp\`{e}re Law (\ref{ampere-phi}) and (\ref{ampere-z}).
As pointed out above,
  $n_e$ is very close to unity for this case,
  so the normalized current density {\bf J} is essentially the negative of the
  normalized electron flow velocity $\bar{\textbf{v}}$.
It is also easy to check that the sign of {\bf J} is consistent with
  the {\bf B} profile.
Since $J_\phi$ is positive and localized,
  it produces an increase of $B_z$ on the $z$-axis.
Asymptotically $B_z$ tends to a constant value $B_\infty$,
  presumably given by the background magnetic field.
At the same time,
  a positive localized $J_z$ produces
  a positive localized $B_\phi$ which tends to an
  asymptotic $1/\rho$ behavior at large $\rho$,
  as consistent with a localized axial current.
For completeness,
  the normalized parallel and 
  perpendicular electron temperatures
  $T_{e \|} $ and $T_{e \perp}$
  are plotted in Fig.~\ref{fig3}~(c) and (d).
They are given by the calculations of the components
  of the normalized electron pressure tensor 
  through Eqs.~(\ref{P-comps}).
We see that they deviate from unity (background) 
  over a region of a length scale close to $l_\psi$.
While the deviation is not large for this case,
  of the order of 10\%,
  it is large enough to be observable if such
  solutions exist in real physical situations.

\begin{figure}
  \centerline{\includegraphics[width=3.3in]{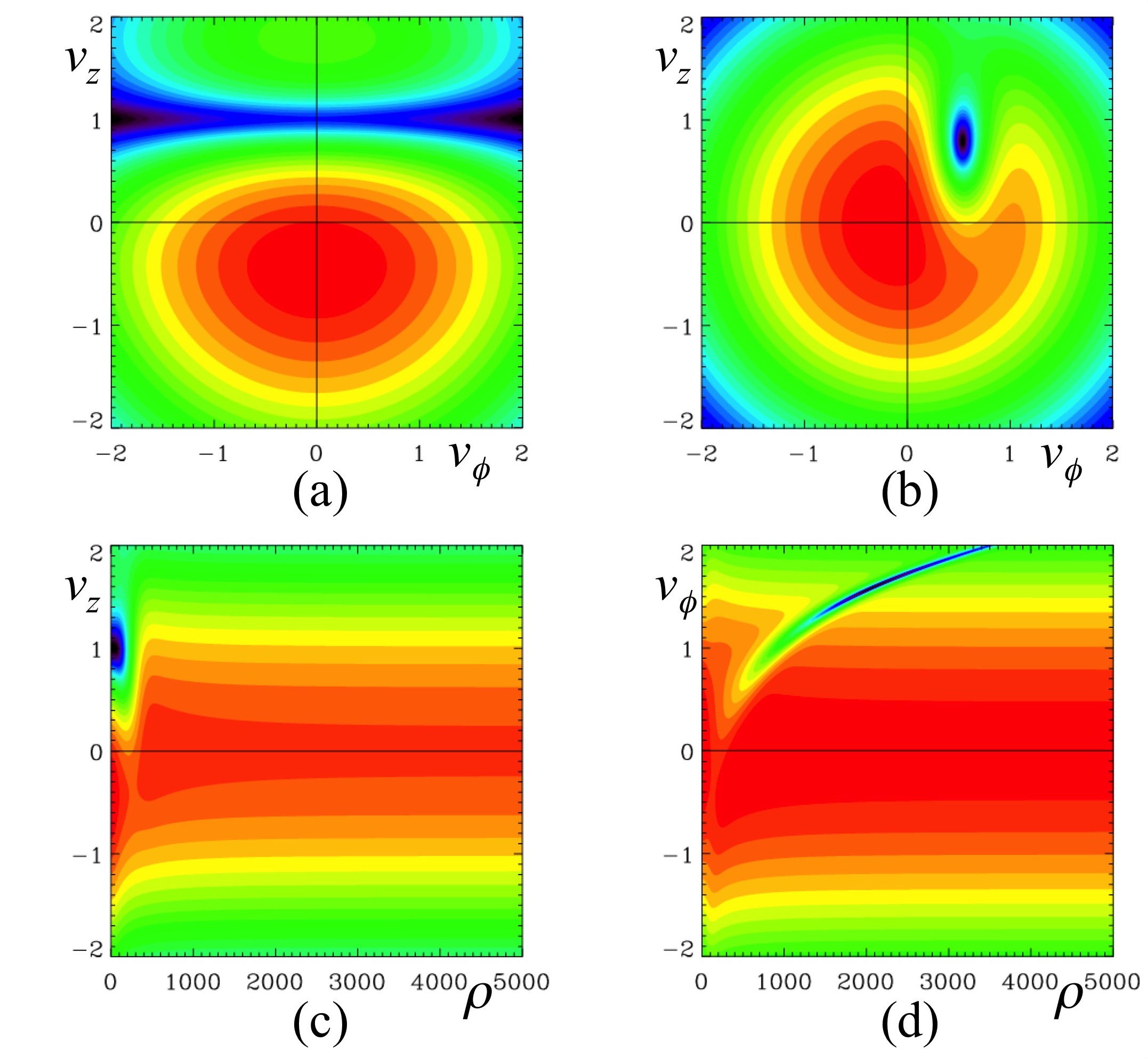}}
\caption{\label{fig4} Color coded contour plots for the cross section of $\ln f_e$
  on the planes of (a) $v_z$ vs. $v_\phi$ at $\rho = v_\rho = 0$, 
  (b) $v_z$ vs. $v_\phi$ at $\rho = 400$, $v_\rho = 0$, 
  (c) $v_z$ vs. $\rho$ at $v_\phi = v_\rho = 0$,
  (d) $v_\phi$ vs. $\rho$ at $v_z = v_\rho = 0$, 
  for the same case as Fig.~\ref{fig1}.}
\end{figure}

With $\psi$, $A_\phi$, and $A_z$ solved from
  Eqs.~(\ref{gauss-rho}), (\ref{ampere-phi}), and (\ref{ampere-z}),
  the corresponding normalized electron distribution $f_e$ can be calculated
  through Eq.~(\ref{f-form}).
For this symmetric solution,
  the phase space of $f_e$ is four-dimensional in $v_\rho$, 
  $v_\phi$, $v_z$, and $\rho$.
However, 
  since the dependence on $v_\rho$ is simply a
  gaussian factor $\exp (-v_\rho^2/2)$,
  non-trivial behavior is in the $v_\phi$, $v_z$, and $\rho$
  space only.
Due to the difficulty in plotting $f_e$ in the 
  full three-dimensional space,
  we will instead plot a few 2D cross sections
  to illustrate some features of the full structure.
Fig.~\ref{fig4}~(a) shows a color coded contour plot
  for the cross section of $\ln f_e$ on the plane of
  $v_z$ vs. $v_\phi$ at $\rho = v_\rho = 0$,
  while Fig.~\ref{fig4}~(b) shows a similar plot 
  but at  $\rho = 400$, 
  where $\psi$ has not decreased to a small level
  with $B_\phi$ and $J_\phi$ near maximum.
In these contour plots,
  we employed a rainbow color coding scheme
  with the dark violet end of the spectrum at 
  the minimum value ($l_{{\rm min}}$) of $\ln f_e$ over the plot area, 
  and the red end of the spectrum at the
  maximum value ($l_{{\rm max}}$) of $\ln f_e$ over the plot area ,
  increasing over 30 contour levels with
  an increment of $\Delta l$ at each level.
For Fig.~\ref{fig4}~(a), $l_{{\rm min}} = -9.34$,
  $l_{{\rm max}} = -2.46$, and $\Delta l = 0.229$,
  while $l_{{\rm min}} = -7.7$,
  $l_{{\rm max}} = -2.68$, and $\Delta l = 0.167$
  for Fig.~\ref{fig4}~(b).
We report these numbers here instead of showing
  a color bar for each plot because such bars would
  be difficult to read or taking up too much space,
  while these numbers would allow the determination of the
  value of $\ln f_e$ at any point if needed.
  
\begin{figure}
  \centerline{\includegraphics[width=3.3in]{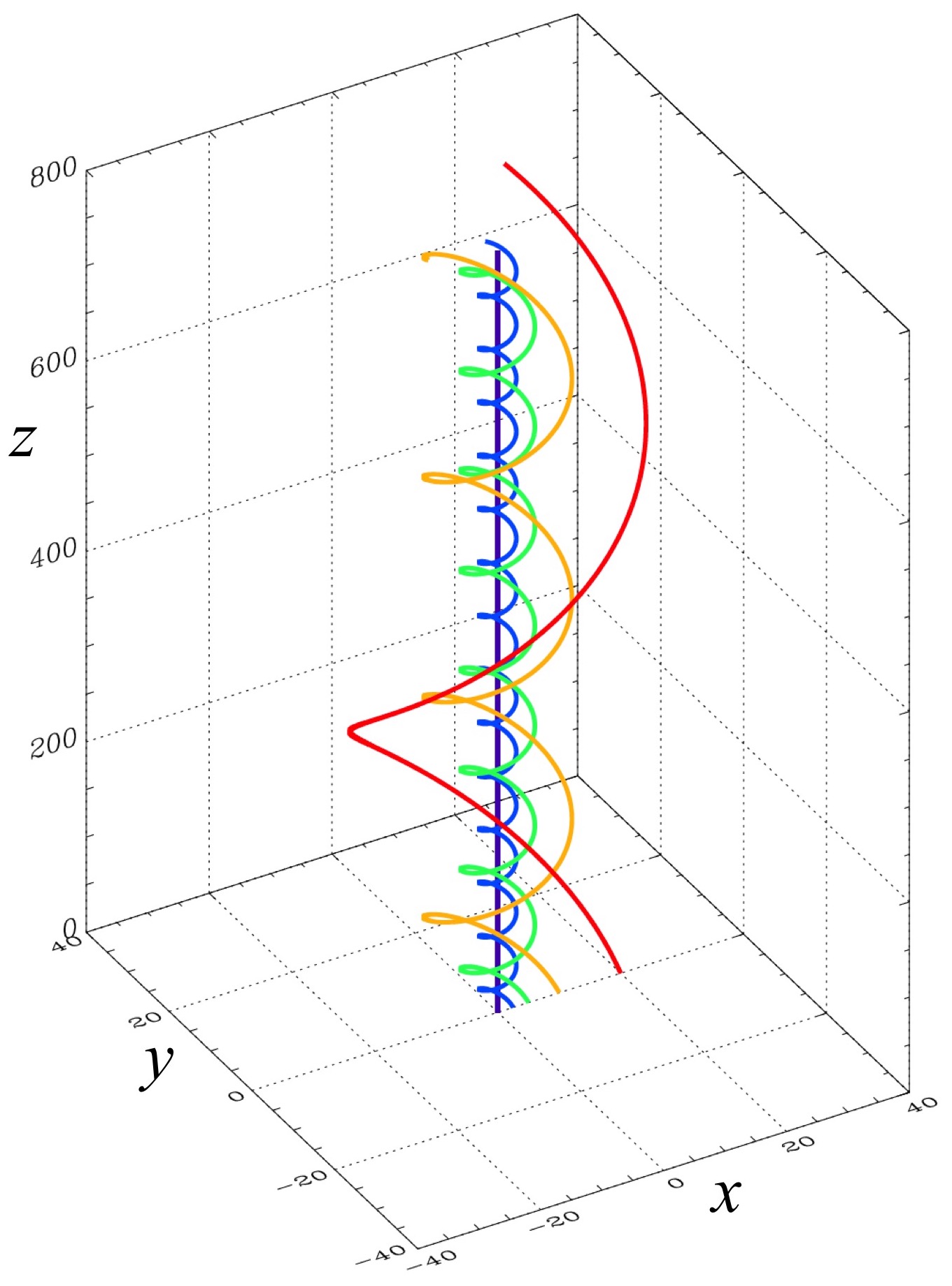}}
\caption{\label{fig5} Five magnetic field lines drawn in
  different colors, going through
  $y = z = 0$, $x = 0$, 2.5, 5, 10, 20 $d_e$,
  which is 200 $\lambda_D$, 
  the unit of $\rho$ in previous plots,
  for the same case as Fig.~\ref{fig1}.}
\end{figure}  
  
From these two plots,
  we see from the shift of the contours with larger values
  that there is an electron flow with a negative
  $z$-component both on the 
  axis ($\rho = 0$), and slightly off axis,
  and a negative $\phi$-component off axis
  but zero on the axis as required by symmetry.
These features are consistent with the plots of 
  $J_\phi$ and $J_z$ in Fig.~\ref{fig3}.
We also see in Fig.~\ref{fig4}~(b) an interesting
  feature of a hole-like structure of electron 
  depletion in the $v_\phi$-$v_z$ space,
  indicating a strong distortion from an isotropic
  distribution as in a Maxwellian.
Fig.~\ref{fig4}~(c) shows the cross section
  on the plane of $v_z$ vs. $\rho$ at $v_\phi = v_\rho = 0$,
  with $l_{{\rm min}} = -8.61$,
  $l_{{\rm max}} = -2.67$, and $\Delta l = 0.198$,
  while Fig.~\ref{fig4}~(d) shows the cross section
  on the plane of $v_\phi$ vs. $\rho$ at $v_z = v_\rho = 0$,
  with $l_{{\rm min}} = -7.35$,
  $l_{{\rm max}} = -2.46$, and $\Delta l = 0.163$.
Electron depletions are seen in these two plots also,
  with a structure more localized in $\rho$ shown
  in Fig.~\ref{fig4}~(c), 
  while the structure extends to much large $\rho$ values.
This is consistent with a narrower profile in $J_z$
  but a more extended profile in $J_\phi$
  shown in Fig.~\ref{fig3}.

With the components of magnetic field $B_z$ and $B_\phi$
  shown in Fig.~\ref{fig2}~(c) and (d),
  it is clear that the magnetic field lines are having a helical
  structure as in a flux rope.
To illustrate this explicitly,
  Fig.~\ref{fig5} shows some magnetic field lines  
  drawn in different colors, 
  going through
  $y = z = 0$, $x = 0$, 2.5, 5, 10, 20 $d_e$
  over a range of positive $z$ values.
We use $d_e$ as the spatial unit for this plot,
  instead of $\lambda_D$ as in previous plots,
  for clarity in the labels,
  and to show that the helical flux rope structure
  has a spatial scale of a few $d_e$.
Since $B_\phi = 0$ at $\rho = 0$ by symmetry,
  the magnetic field on the $z$-axis is going straight
  along the axis.
Similarly,
  since $B_z \rightarrow B_\infty$ 
  while $B_\phi \rightarrow C_1/\rho$
  as $\rho \rightarrow \infty$,
  the magnetic field lines also tend to
  uniform field along the $z$-direction asymptotically.
Therefore the helical flux rope structure
  with larger $B_\phi$ is localized
  to a range of $\rho$ approximately
  between 1 $d_e$ and 10 $d_e$.

\begin{figure}
  \centerline{\includegraphics[width=3.3in]{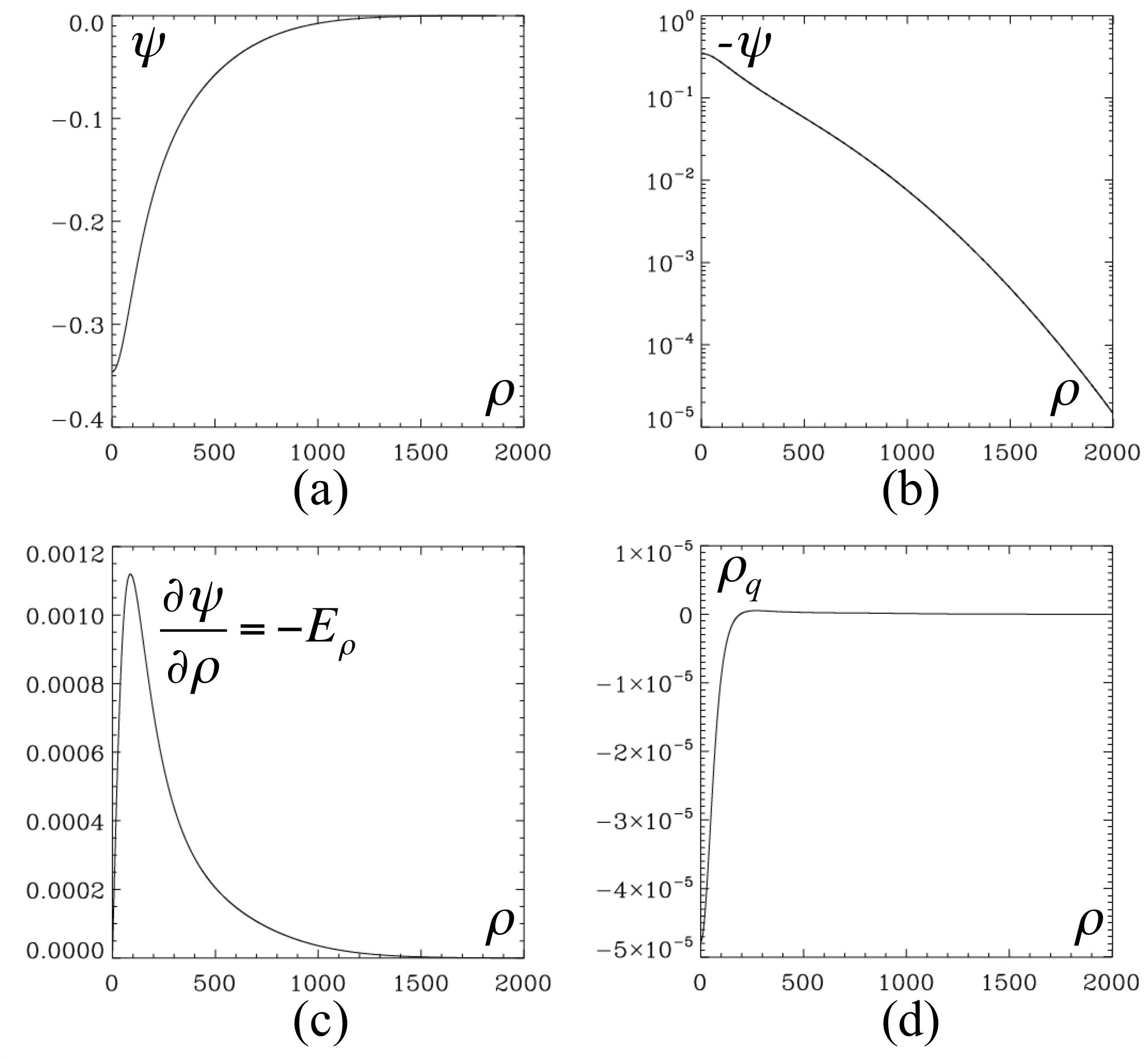}}
\caption{\label{fig6} (a) $\psi$, (b) $-\psi$ in semi-log scales, 
  (c) $d\psi/d\rho$, (d) $\rho_q$ as functions of $\rho$, 
  for a case with $h = -1$, $k = 1 \times 10^{-5}$,
  $\xi = 1$, $\beta_e = 0.005$, $A_z(0) = 1$,
  and $B_z(0) = 0.00293$.}
\end{figure}

\begin{figure}
  \centerline{\includegraphics[width=3.3in]{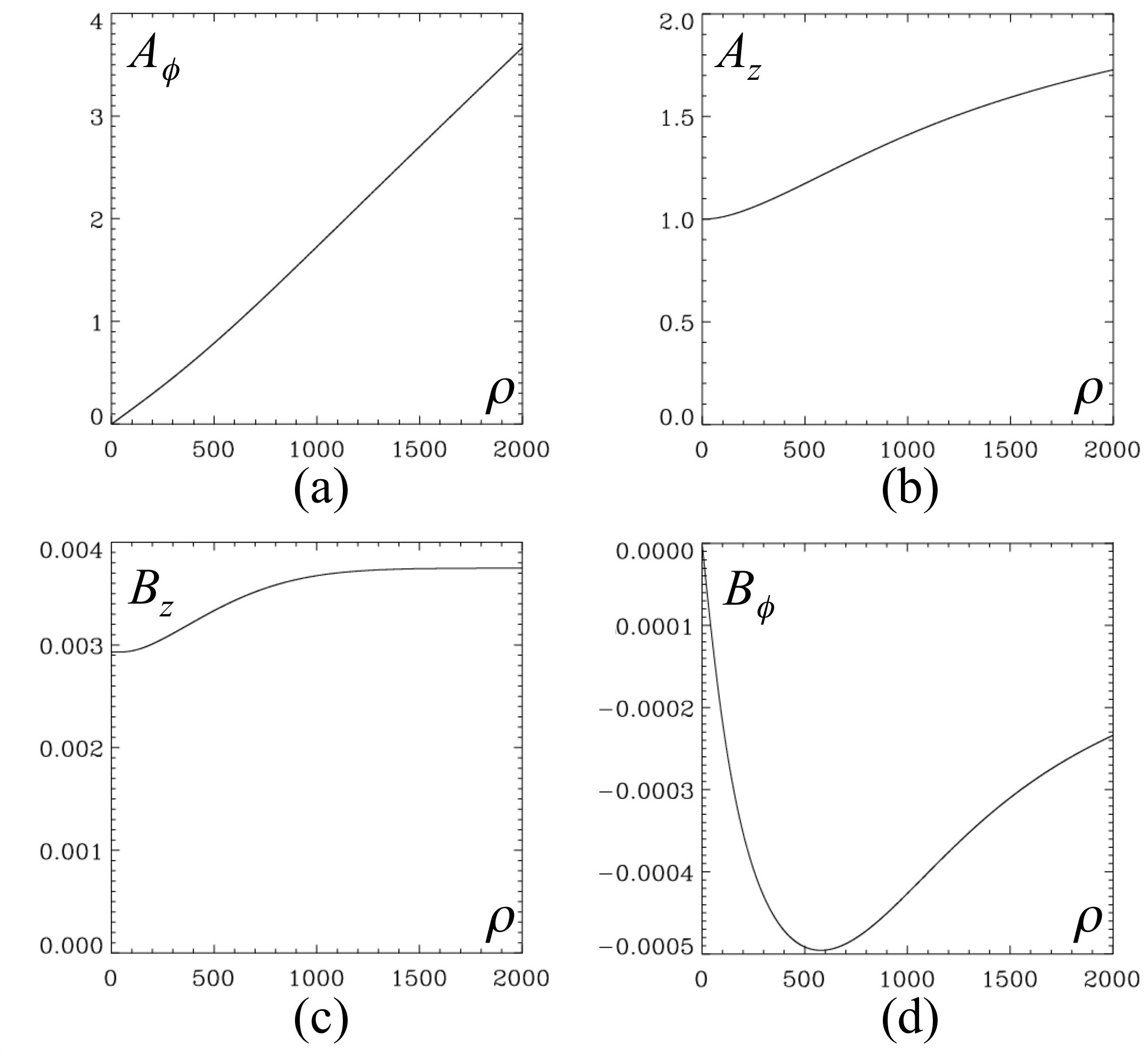}}
\caption{\label{fig7} (a) $A_\phi$, (b) $A_z$, 
  (c) $B_z$, (d) $B_\phi$ as functions of $\rho$, 
  for the same case as Fig.~\ref{fig6}.}
\end{figure}

\begin{figure}
  \centerline{\includegraphics[width=3.3in]{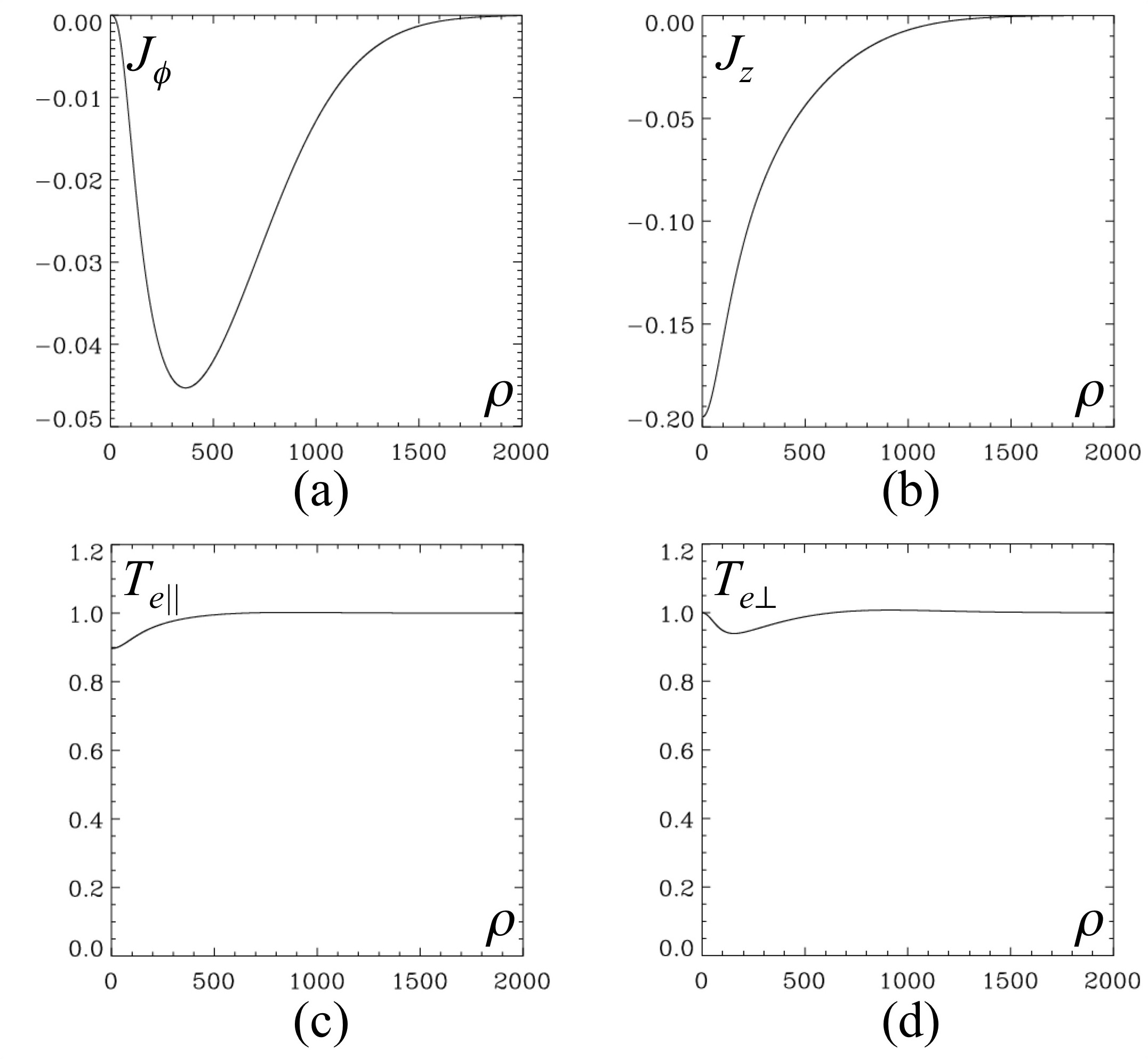}}
\caption{\label{fig8} (a) $J_\phi$, (b) $J_z$, 
  (c) $T_{e\parallel}$, (d) $T_{e\perp}$ as functions of $\rho$, 
  for the same case as Fig.~\ref{fig6}.}
\end{figure}

\begin{figure}
  \centerline{\includegraphics[width=3.3in]{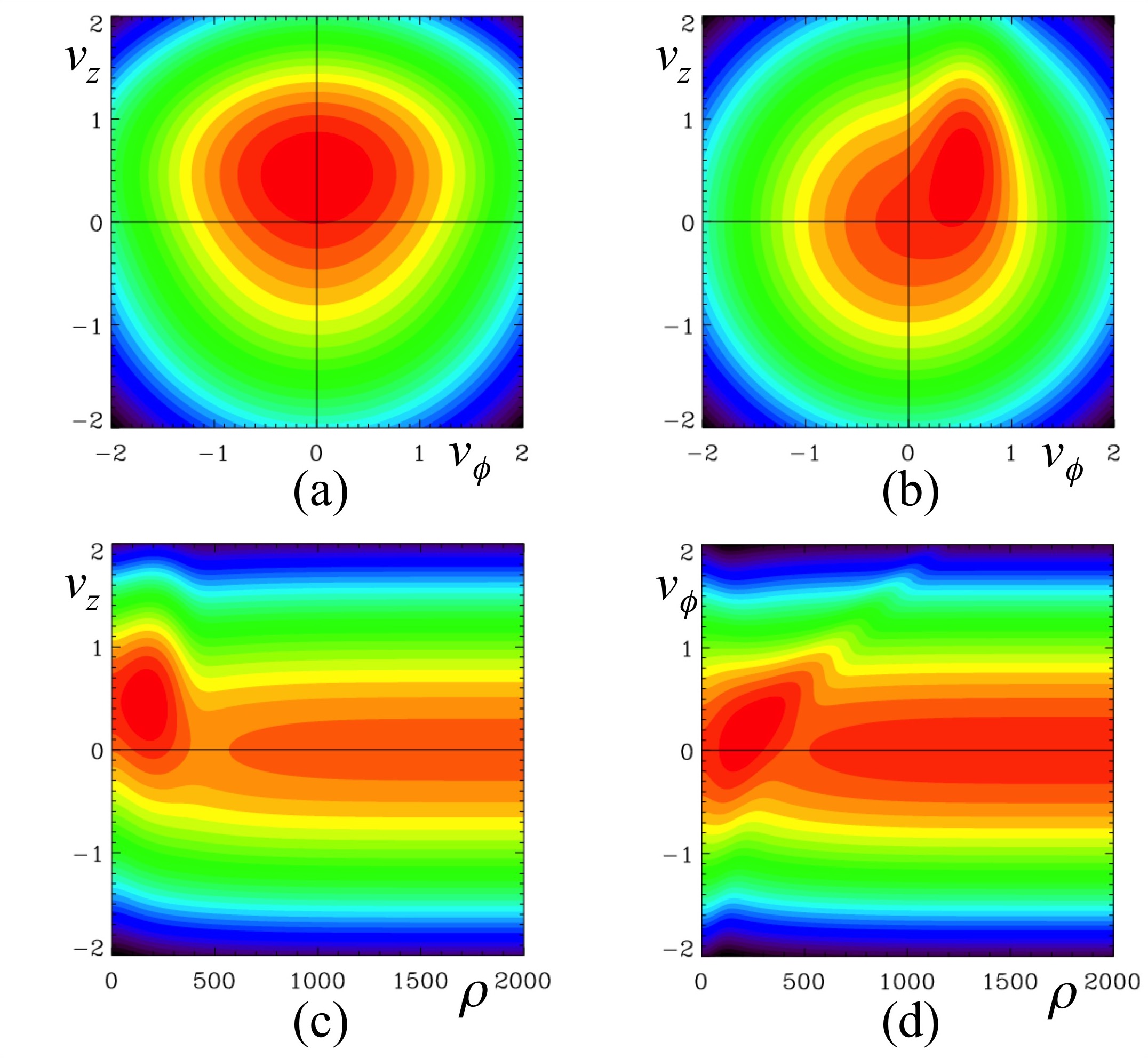}}
\caption{\label{fig9} Color coded contour plots for the cross section of $\ln f_e$
  on the planes of (a) $v_z$ vs. $v_\phi$ at $\rho = v_\rho = 0$, 
  (b) $v_z$ vs. $v_\phi$ at $\rho = 400$, $v_\rho = 0$, 
  (c) $v_z$ vs. $\rho$ at $v_\phi = v_\rho = 0$,
  (d) $v_\phi$ vs. $\rho$ at $v_z = v_\rho = 0$, 
  for the same case as Fig.~\ref{fig6}.}
\end{figure}

For the second case,
  Figs.~\ref{fig6} to \ref{fig9} show plots corresponding
  to Figs.~\ref{fig1} to \ref{fig4} for a case with 
  $h = -1$, $k = 1 \times 10^{-5}$,
  $\xi = 1$, $\beta_e = 0.005$, $A_z(0) = 1$,
  and $B_z(0) = 0.00293$.
We see from these plots that many features for this
  case are very similar to the first case
  except with opposite sign.
The spatial scales of structures are actually similar
  but we plot in Figs.~\ref{fig6} to \ref{fig9} using a range of $\rho$ from
  0 to $2000 \lambda_D$ instead of $5000 \lambda_D$
  as in Figs.~\ref{fig1} to \ref{fig4} to show more clearly 
  features near the axis for comparison.
Electric potential $\psi$ and electric field $E_\rho$,
  as well as charge density $\rho_q$,
  shown in Fig.~\ref{fig6} are with signs negative
  to corresponding quantities shown in Fig.~\ref{fig1},
  although not with exactly the same magnitudes.
$B_z$ shown in Fig.~\ref{fig7}~(c) still has positive sign
  by choice but is having a dip in value on and near the
  $\rho = 0$ axis,
  opposite to an increase shown in Fig.~\ref{fig2}~(c).
$B_\phi$ for this case shown in Fig.~\ref{fig7}~(d)
  is now pointing to the negative direction,
  opposite to the first case shown in in Fig.~\ref{fig2}~(d).

These opposite features in the magnetic field is indeed
  due to current density components $J_\phi$
  and $J_z$ having negative sign, 
  as shown in Fig.~\ref{fig8}~(a) and (b),
  opposite to corresponding plots shown in 
  Fig.~\ref{fig3}~(a) and (b) for the first case.
The normalized parallel and 
  perpendicular electron temperatures
  $T_{e \|} $ and $T_{e \perp}$
  plotted in Fig.~\ref{fig8}~(c) and (d)
  are also having different profiles than 
  those in Fig.~\ref{fig3}~(c) and (d),
  although not exactly opposite.

Cross section plots of $\ln f_e$ for this case shown in Fig.~\ref{fig9}
  are also very different from corresponding plots
  shown in Fig.~\ref{fig4}.
We first list color code values to read these plots:
  $l_{{\rm min}} = -7.1$,
  $l_{{\rm max}} = -2.64$, and $\Delta l = 0.148$
  for Fig.~\ref{fig9}~(a);
  $l_{{\rm min}} = -6.84$,
  $l_{{\rm max}} = -2.61$, and $\Delta l = 0.141$
  for Fig.~\ref{fig9}~(b);
  $l_{{\rm min}} = -4.98$,
  $l_{{\rm max}} = -2.65$, and $\Delta l = 0.0777$
  for Fig.~\ref{fig9}~(c);
  $l_{{\rm min}} = -6.84$,
  $l_{{\rm max}} = -2.55$, and $\Delta l = 0.143$
  for Fig.~\ref{fig9}~(d).
This case with $h = -1$ means that now electrons
  are added to the Boltzmann distribution,
  rather than taking away in the first case with $h = 0.99$,
  as can be seen from the from of the distribution 
  specified in Eq.~(\ref{f-form}).
Therefore we see from Fig.~\ref{fig9} increases of electrons in
  regions corresponding to electron depletions shown 
  in Fig.~\ref{fig4}.
Similar to the discussion on Fig.~\ref{fig4},
  the shift of the electron distribution seen in Fig.~\ref{fig9}
  is consistent with the sign of the current density 
  plotted in Fig.~\ref{fig8}.

\section{DISCUSSION AND CONCLUSION \label{conclu}}

In this paper, 
  we have presented a general theory in constructing
  localized 2D BGK modes,
  which are exact solutions of the 
  Vlasov-Poisson-Amp\`{e}re system of equations,
  with electron distribution functions
  depending on three conserved quantities: energy, 
  canonical angular momentum, 
  and canonical momentum along the symmetric axis.
These 2D solutions generally have a magnetic field 
  in both the axial and azimuthal directions and thus
  a structure in the form of a flux rope in the kinetic scales.
We emphasize that such flux rope solutions are solved
  from the kinetic theory (Vlasov equation) and thus are
  very different from flux rope solutions in MHD.

To show the existence of solutions based on this theory,
  and to illustrate some general features in solutions,
  we have presented two solutions found numerically
  based on a special form of solution specified in
  Eq.~(\ref{f-form}). 
This form of solutions depends on six parameters.
The two examples are chosen with only one parameter
  being different: $h = 0.99$ for the first case,
  which can be described physically as taking away electrons
  so that electron holes form in the electron distribution,
  while $h = -1$ for the second case,
  which has electrons added instead. 
While magnetic flux rope structures exist in both cases,
  electric potential and electric fields,
  as well as current density and magnetic field perturbations
  are having opposite signs,
  most noticeably an increase of magnetic field on the axis
  for the first case,
  but a decrease for the second case.
Correspondingly there is a decrease of the electron density
  $n_e$ on the axis for the first case,
  but an increase for the second case,
  although $n_e \approx 1$ with our choices of parameters
  in our examples.
  
The second case with a negative potential $\psi$ deserves
  some more discussion.
Such localized electrostatic structure of negative electric 
  potential produced by electron dynamics with a uniform
  ion background does not exist in the conventional 1D
  BGK mode solution,
  since the existent of a self-consistent solution in that 
  case depends on having trapping electrons with
  a different form of distribution from the Boltzmann distribution
  of passing electrons. 
However, 
  a localized negative electric potential cannot trap electrons
  and thus all electrons are having the same Boltzmann distribution
  which cannot support a self-consistent negative potential structure.
The fact that a solution with a negative potential exists as a 2D
  BGK mode is interesting in itself.
Moreover, 
  there is an indication that a negative potential solution might be
  easier to exist than a positive one
  in the sense that the range of $h$ for the positive case in our
  form of solutions is $0 < h < 1$,
  while it is $-\infty < h < 0$ for the negative potential case.
Physically it means that while there is a limit in taking electrons
  away from a Boltzmann distribution since the most one can
  do is to decrease it to zero, 
  there is no limit in adding electrons in the negative
  potential case.
  
We again emphasize that the two examples presented in this
  paper are simply two choices of parameters out of a large
  six-dimensional parameter space,
  and thus there can be solutions with very different properties
  from these two examples.
Moreover, 
  as have been pointed out above,
  our form of the electron distribution in Eq.~(\ref{f-form}) can
  be easily generalized to forms with many more parameters.
Due to the non-linear nature of such solutions,
  such generalizations most likely will produce different solutions.
Adding to the complication is the ion dynamics,
  which is ignored in this paper with the assumption of a uniform
  ion background.
There is one more possible generalization in extending 
  the form of the electron distribution from a smooth and analytic
  ones such as in Eq.~(\ref{f-form}) to possible discontinuous 
  ones such as in 1D BGK mode theory after taking into 
  account of the possibility of trapping electrons having a
  different distribution.
The physics of trapping (in transverse directions only) is ignored
  in this paper but it can indeed be included in a more general
  case.

With the possibility of the existent of 2D BGK mode solutions 
  over a very large parameter space,
  it is interesting to see whether such solutions can be compared
  with observations in space or experiments.
However, 
  this also depends on the stability of such solutions,
  as well as the existent of mechanisms that can generate
  such solutions.
In this paper,
  we have only considered the existent of steady-state 
  solutions,
  without considering whether such solutions are stable
  and how such solutions can be formed.
While these are interesting and important topics for 
  further research,
  our solutions can also be used to compare with
  large-scale numerical simulations using either
  the Vlasov equation or PIC method,
  in which flux ropes in kinetic scales have been
  observed.



%
%

%

\begin{acknowledgments}
This work is partially supported by a National Science Foundation grant PHY-1004357.
\end{acknowledgments}

\bibliography{KFR}

\end{document}